\documentclass[]{nature}
\makeatletter\if@twocolumn\PassOptionsToPackage{switch}{lineno}\else\fi\makeatother

\def\CVS{CsV$_3$Sb$_5$}
\def\KVS{KV$_3$Sb$_5$}
\def\RVS{RbV$_3$Sb$_5$}
\def\AVS{$A$V$_3$Sb$_5$}
\def\cm{cm$^{-1}$}
\def\Tc{$T_{\rm CDW}$}

\def\wp{$\omega_{\mathrm{p}}$}
\usepackage{tabulary,graphicx,amsmath,amsfonts,amssymb}
\usepackage[utf8x]{inputenc}
\usepackage[T1]{fontenc}

\makeatletter

\renewenvironment{table*}
        {\@dblfloat{table}}
           {\end@dblfloat}

\makeatother

\usepackage{url,multirow,morefloats,floatflt,cancel,tfrupee}
\makeatletter

\AtBeginDocument{\@ifpackageloaded{textcomp}{}{\usepackage{textcomp}}}
\makeatother
\usepackage{colortbl}
\usepackage{xcolor}
\usepackage{pifont}
\usepackage[nointegrals]{wasysym}
\makeatletter

\def\mcWidth#1{\csname TY@F#1\endcsname+\tabcolsep}

\def\cAlignHack{\rightskip\@flushglue\leftskip\@flushglue\parindent\z@\parfillskip\z@skip}
\def\rAlignHack{\rightskip\z@skip\leftskip\@flushglue \parindent\z@\parfillskip\z@skip}

\@ifundefined{etal}{}{}

\usepackage{ifxetex}
\ifxetex\else\if@twocolumn\@ifpackageloaded{stfloats}{}{\usepackage{dblfloatfix}}\fi\fi

\AtBeginDocument{
\expandafter\ifx\csname eqalign\endcsname\relax
\def\eqalign#1{\null\vcenter{\def\\{\cr}\openup\jot\m@th
  \ialign{\strut$\displaystyle{##}$\hfil&$\displaystyle{{}##}$\hfil
      \crcr#1\crcr}}\,}
\fi
}

\AtBeginDocument{%
  \@ifpackageloaded{endfloat}%
   {\renewcommand\efloat@iwrite[1]{\immediate\expandafter\protected@write\csname efloat@post#1\endcsname{}}}{\newif\ifefloat@tables}%
}%

\def\BreakURLText#1{\@tfor\brk@tempa:=#1\do{\brk@tempa\hskip0pt}}
\let\lt=<
\let\gt=>
\def\processVert{\ifmmode|\else\textbar\fi}

\@ifundefined{subparagraph}{
\def\subparagraph{\@startsection{paragraph}{5}{2\parindent}{0ex plus 0.1ex minus 0.1ex}%
{0ex}{\normalfont\small\itshape}}%
}{}

\newcommand\role[1]{\unskip}
\newcommand\aucollab[1]{\unskip}
  
\@ifundefined{tsGraphicsScaleX}{\gdef\tsGraphicsScaleX{1}}{}
\@ifundefined{tsGraphicsScaleY}{\gdef\tsGraphicsScaleY{.9}}{}
\def\checkGraphicsWidth{\ifdim\Gin@nat@width>\linewidth
	\tsGraphicsScaleX\linewidth\else\Gin@nat@width\fi}

\def\checkGraphicsHeight{\ifdim\Gin@nat@height>.9\textheight
	\tsGraphicsScaleY\textheight\else\Gin@nat@height\fi}

\def\fixFloatSize#1{}
\let\ts@includegraphics\includegraphics

\def\inlinegraphic[#1]#2{{\edef\@tempa{#1}\edef\baseline@shift{\ifx\@tempa\@empty0\else#1\fi}\edef\tempZ{\the\numexpr(\numexpr(\baseline@shift*\f@size/100))}\protect\raisebox{\tempZ pt}{\ts@includegraphics{#2}}}}

\AtBeginDocument{\def\includegraphics{\@ifnextchar[{\ts@includegraphics}{\ts@includegraphics[width=\checkGraphicsWidth,height=\checkGraphicsHeight,keepaspectratio]}}}

\DeclareMathAlphabet{\mathpzc}{OT1}{pzc}{m}{it}

\def\URL#1#2{\@ifundefined{href}{#2}{\href{#1}{#2}}}

\def\UrlOrds{\do\*\do\-\do\~\do\'\do\"\do\-}%
\g@addto@macro{\UrlBreaks}{\UrlOrds}

\edef\fntEncoding{\f@encoding}

\makeatother

\newif\ifmultipleabstract\multipleabstractfalse%
\usepackage[nomarkers,tablesfirst,nolists]{endfloat}
\usepackage{lineno}[displaymath,mathlines]

\def\fixFloatSize#1{}
\begin{document}

\title{Pressure evolution of electron dynamics in the superconducting kagome
metal CsV$_3$Sb$_5$}

\author{Maxim Wenzel$^{1,}$\thanks{E-mail: maxim.wenzel@pi1.physik.uni-stuttgart.de},~Alexander A. Tsirlin$^{2,3,}$\thanks{E-mail: altsirlin@gmail.com},~Francesco Capitani$^{4}$,~Yuk T. Chan$^{1}$,~Brenden R. Ortiz$^{5,6}$,~Stephen D. Wilson$^{5}$,~Martin Dressel$^{1}$,~Ece Uykur$^{1,7,}$\thanks{E-mail: e.uykur@hzdr.de}
}

\maketitle 
\begin{affiliations}
  \item 
    1. Physikalisches Institut, Universität Stuttgart, D-70569 Stuttgart, Germany
    \item 
    Felix Bloch Institute for Solid-State Physics, Leipzig University, 04103 Leipzig, Germany
    \item
    Experimental Physics VI, Center for Electronic Correlations and Magnetism, Institute of Physics, University of Augsburg, 86135 Augsburg, Germany
    \item
    Synchrotron SOLEIL, L’Orme des Merisiers, Saint-Aubin, 91192 Gif-sur-Yvette, France
    \item
    Materials Department and California Nanosystems Institute,
University of California Santa Barbara, Santa Barbara, CA, 93106, United States
\item
Materials Department, University of California Santa Barbara, Santa Barbara, CA, 93106, United States
\item
Helmholtz-Zentrum Dresden-Rossendorf, Institute of Ion Beam Physics and Materials Research, 01328 Dresden, Germany
    
    \end{affiliations}\def\keywordstitle{Keywords}

\begin{abstract}
The coexistence of the charge-density wave (CDW) and superconducting phases and their tunability under external pressure remains one of the key points in understanding the electronic structure of $A$V$_3$Sb$_5$ ($A$~=~K, Rb, Cs) kagome metals. Here, we employ synchrotron-based infrared spectroscopy assisted by density-functional calculations to study the pressure evolution of the electronic structure at room temperature up to 17~GPa experimentally. The optical spectrum of CsV$_3$Sb$_5$ is characterized by the presence of localized carriers seen as a broad peak at finite frequencies in addition to the conventional metallic Drude response. The non-monotonic pressure dependence of this low-energy peak reflects the re-entrant behavior of superconductivity and may be interpreted in terms of electron-phonon coupling, varying with the growth and shrinkage of the Fermi surface (FS) under pressure. Moreover, drastic modifications in the low-energy interband absorptions are observed upon the suppression of CDW. These changes are related to the upward shift of the Sb2~$p_x + p_y$ band that eliminates part of the FS around the $M$-point, whereas band saddle points do not move significantly. These observations shed new light on the mixed electronic and lattice origin of the CDW in \CVS.
\end{abstract}\def\keywordstitle{Keywords}


\section*{Introduction}

Discovered in 2019\cite{Ortiz2019}, the non-magnetic kagome metal series \AVS\ ($A$ = K, Rb, Cs) provides an exciting playground for studying a plethora of fascinating electronic phenomena, including the interplay between a charge-density wave (CDW) phase and superconductivity. In \CVS, the CDW instability forms below T$_{\mathrm{CDW}} = 94$~K\cite{Ortiz2020} (102~K for \RVS\cite{Yin2021} and 78~K for \KVS\cite{Ortiz2019}), while superconductivity (SC) is observed below $T_{\mathrm{c}}=2.5$~K\cite{Ortiz2020} (0.92~K for \RVS\cite{Yin2021} and 0.93~K for \KVS\cite{Ortiz2021}).

Recently, several electrical transport, magnetic susceptibility, NMR, XRD, and $\mu$SR studies demonstrated the tunability of both orders in the \AVS\ series under external pressure\cite{Chen2021, Wang2021, Yu2021, Zhao2021, Zhu2022, Li2022, Chen2021b, Guguchia2022, Wang2021b, Du2021, Zhang2021, Zheng2022, Gupta2022}. At low pressures ($p < 2$~GPa), superconductivity in \CVS\ exhibits a double-dome feature, comprising an enhancement of $T_{\mathrm{c}}$ to $\sim$~7~K at $p_{\rm 1} \sim 0.6$~GPa, and a second peak in $T_{\mathrm{c}}$ at $p_{\rm 2} \sim 2$~GPa with $T_{\mathrm{c}} \sim 8$~K, whereas the superconducting gap evolves almost monotonically\cite{Wen2023}. Simultaneously, \Tc\ is gradually suppressed and vanishes at $p_{\rm 2}$\cite{Chen2021, Yu2021, Li2022}. Experimental studies revealed a non-trivial evolution of CDW across $p_{\rm 1}$, suggesting that changes in the nature of CDW may cause the double-dome behavior of superconductivity\cite{Zheng2022, Feng2023, Li2022}. For $p > 2$~GPa, superconductivity is gradually suppressed and vanishes around 9~GPa. Surprisingly, at higher pressures ($p > 12$~GPa), superconductivity re-emerges, with this second SC phase persisting up to at least 100~GPa \cite{Zhao2021, Chen2021b, Zhang2021}. This re-entrant behavior, along with the presence of a quantum critical point (QCP) at 2~GPa, has sparked considerable debate on the superconducting pairing mechanism. Some studies propose electronic correlations at ambient pressure and CDW fluctuations around 2~GPa to be the main driving force behind the formation of Cooper pairs\cite{Neupert2022, Wu2021, Tazai2022}. On the other hand, first-principle calculations revealed essential changes in electron-phonon (e-ph) coupling around the QCP, suggesting a conventional phonon-mediated mechanism\cite{Zhang2021b, Si2022, Wang2022, Wang2023}. 

The CDW state in \CVS\ is unique within the \AVS\ series as both a 2~$\times$~2~$\times$~2 and 2~$\times$~2~$\times$~4 order comprising stacked 2~$\times$~2 in-plane star-of-David and trihexagonal kagome superlattices are possible at ambient pressure\cite{Ortiz2021a, Hu2022, Kang2022}. Moreover, the experimentally observed CDW state features multiple anisotropic energy gaps \cite{Wang2021a, Nakayama2021}, large anomalous Hall effect \cite{Yu2022, Yu2021a}, intrinsic chirality \cite{Yu2021a, Wang2021c, Guo2022}, and a temperature-driven re-arrangement of the order parameters including an electronic nematic phase below $\sim$~35~K\cite{Nie2022, Zheng2022, Stahl2022}. Some studies discuss electron-phonon coupling as the driving force behind the CDW instability in \AVS\cite{Si2022, Xie2022, Luo2022}, whereas others point towards a more complex electronic origin \cite{Tan2021, Li2021}. It has been shown that band saddle points (van Hove singularities) close to the Fermi energy lead to a CDW instability in kagome metals\cite{Kiesel2013, Park2021, Denner2021}, making the pressure evolution of the saddle points a potentially important ingredient for understanding the pressure-phase diagram of \CVS.

Linking the pressure-induced changes in the electronic properties to modifications in the electronic band structure has been the subject of several high-pressure XRD studies, along with density functional theory calculations\cite{Tsirlin2022, Tsirlin2022b, Chen2021b, Wang2022, LaBollita2021, Yu2022b, Si2022, Zhang2021b}. The experimental confirmation of the proposed changes in the electronic band structure, however, remains lacking because the powerful ambient-pressure probes, such as ARPES and STM, are incompatible with pressure environment. Hence, we utilize high-pressure Fourier-transform infrared spectroscopy at room temperature to experimentally probe the pressure-induced modifications in the electronic structure. Assessing these changes in the normal state is crucial for understanding the changes in low-temperature instabilities of the system, because those instabilities are driven by the evolution of the electronic structure in the normal state.

\begin{figure*}
\centering 
\includegraphics{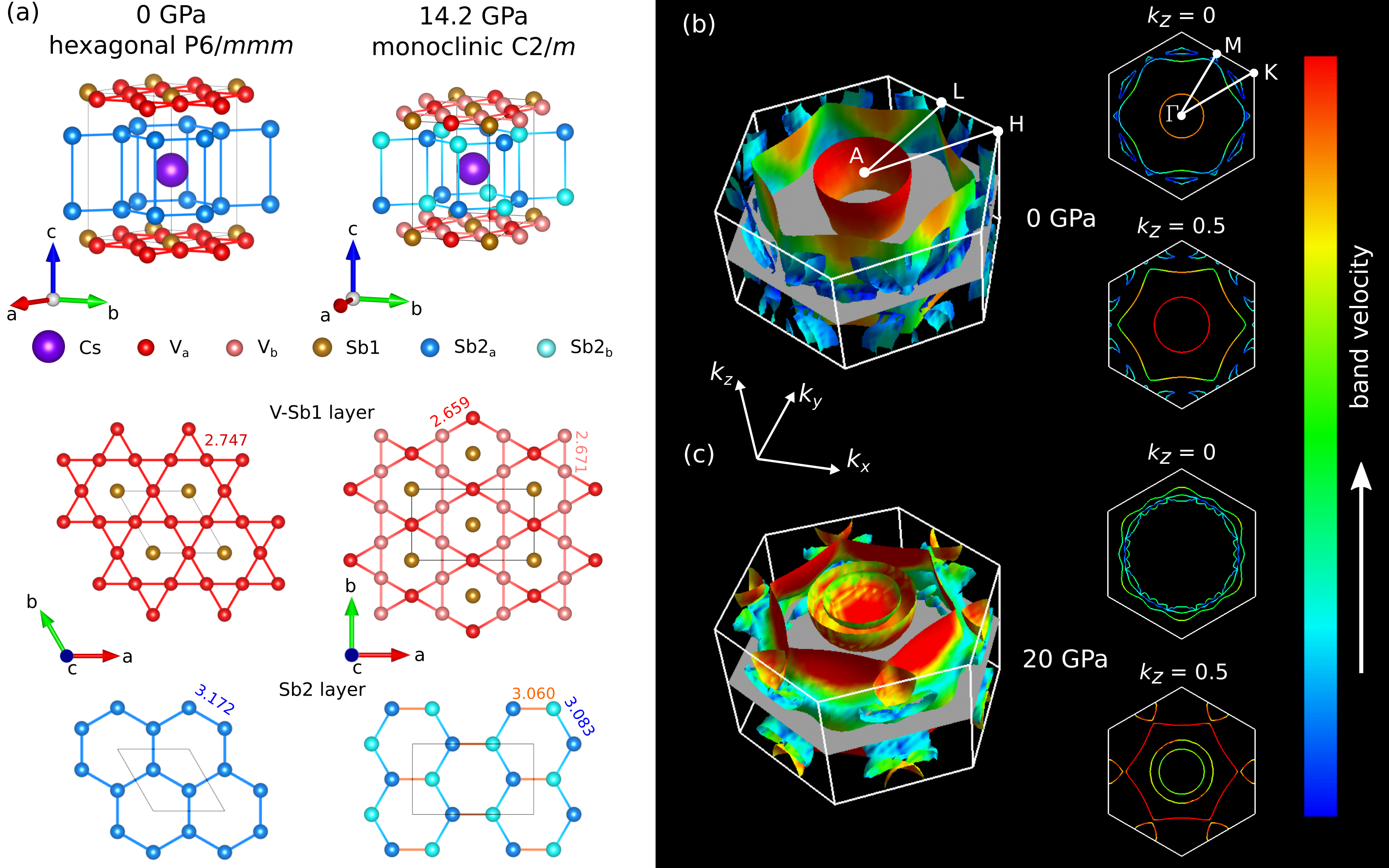}
\caption{\textbf{Pressure evolution of the crystal structure and Fermi surface.} \textbf{a} Hexagonal (left) and monoclinic (right) crystal structures of \CVS\ as determined by XRD measurements at ambient pressure\cite{Ortiz2019} and 14.2~GPa under non-hydrostatic conditions\cite{Tsirlin2022b}, respectively, visualized by \texttt{VESTA}\cite{Momma2008}. \textbf{b}~and~\textbf{c} Representative Fermi surfaces, at 0 and 20~GPa, respectively, created with \texttt{FermiSurfer}\cite{Kawamura2019}. Hexagonal structures\cite{Tsirlin2022} were used for the calculations. The color code represents the band velocity. }
\label{Fig1}
\end{figure*}

As most high-pressure transport measurements are performed using silicone or Daphne oil as pressure transmitting medium, increasing pressure leads to increasing non-hydrostaticity\cite{Klotz2009, Perez2011}. It has been shown that under such conditions, \CVS\ transforms from hexagonal to monoclinic symmetry above $\sim$~10~GPa\cite{Tsirlin2022b} as sketched in Fig.~\ref{Fig1}~(a). The current reflectivity study has been performed with the quasi-hydrostatic CsI\cite{Celeste2019} as a pressure transmitting medium, ensuring a good sample-diamond interface, which is crucial for reliable measurements. Hence, here, a similar phase transition can also be envisaged. However, as discussed previously\cite{Tsirlin2022b}, this monoclinic distortion affects the band structure only marginally. 

Our results reveal a drastic change in the interband absorption around $p_{\rm 2}$ where the CDW is suppressed. Using \textit{ab initio} calculations of the band structure and optical conductivity, we show that these changes can be understood in terms of the upward shift of the Sb2~$p_x + p_y$ band and the shrinkage of the Fermi surface around the $M$-point. We further demonstrate that the spectral weight due to localized carriers is strongly suppressed around $p_{\rm 2}$, suggesting the reduction in the e-ph coupling. This spectral weight is partially restored above 10~GPa, where re-entrant superconductivity is observed. Our data suggest that changes in the e-ph coupling caused by the reconstruction of the FS should be crucial for the pressure evolution of electron dynamics in \CVS.

\section*{Results}
\textbf{Intraband contributions.} The decomposed real part of the in-plane optical conductivity at selected pressure points is given in Figs.~\ref{Fig2}~(a-d) (see Supplementary Note 2 for details on the fitting process). While the high-energy contributions ($\omega$~$>$~4000~\cm) are not notably affected by the applied pressure, major changes occur at low energies ($\omega$~$<$~2000~\cm). The optical spectrum at ambient pressure ($p~=~0$~GPa), reproduced from ref.~\cite{Uykur2021b}, is characterized by several interband absorptions and two clearly separated intraband contributions (i) a sharp Drude peak due to free charge carriers and (ii) an additional broad peak centered at finite frequencies corresponding to the response of localized carriers. Both of these intraband features are found to be highly sensitive to the applied pressure, as illustrated in Figs.~\ref{Fig2} (e) and (f). Already at moderate pressures ($p~<~2.3$~GPa), the intensity of the localization peak is drastically reduced and continues to decrease up to $\sim$~9~GPa. Upon further increasing the pressure, the localization peak becomes more pronounced again. On the other hand, the opposite trend is observed for the Drude contribution, signaling an interplay between localized and free charge carriers.

\begin{figure*}
\centering 
\includegraphics{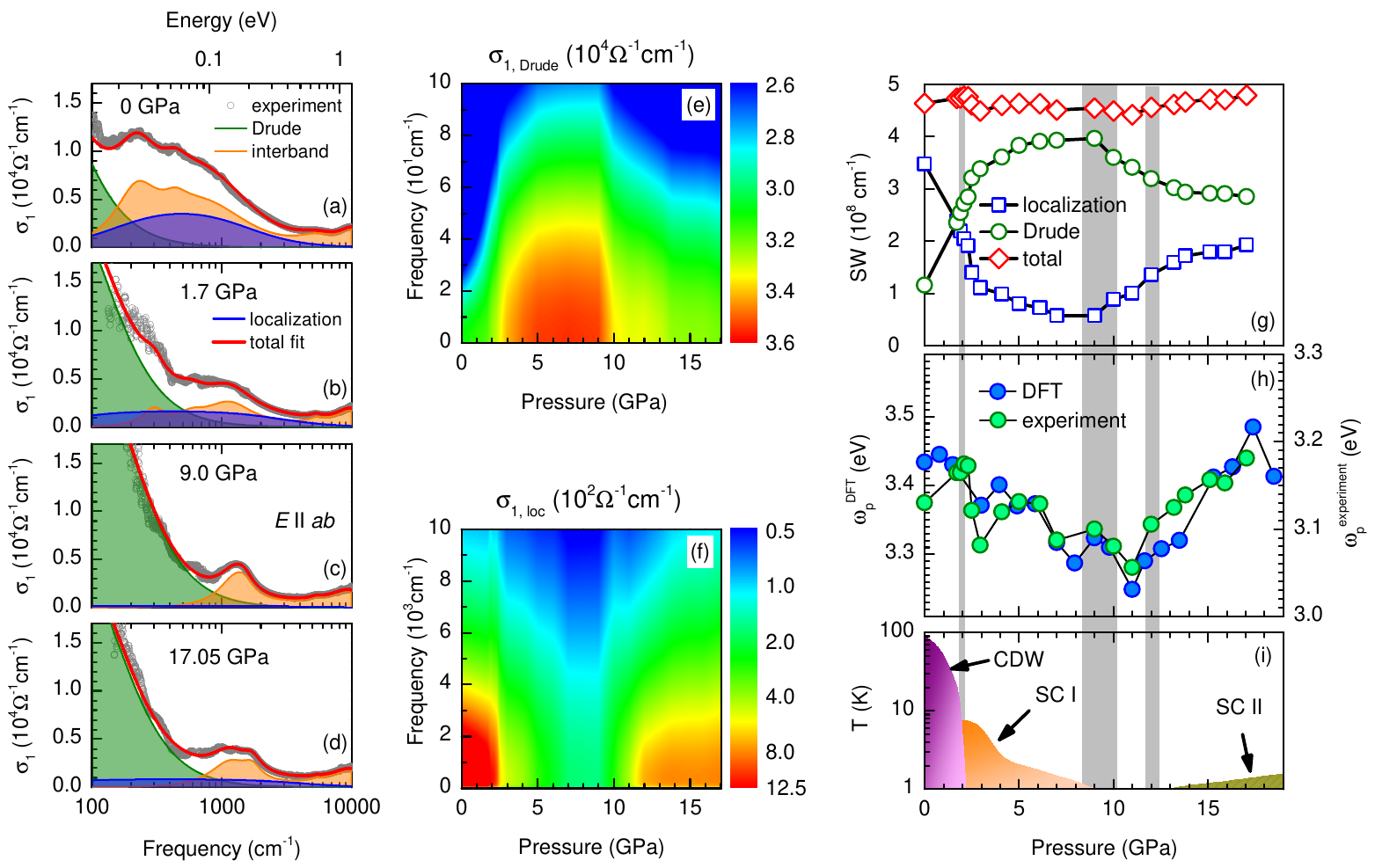}
\caption{\textbf{Decomposition of optical conductivity and intraband spectral weight analysis.} \textbf{a-d} Decomposed optical conductivity at several pressure points consisting of a Drude peak (green), a localization peak (blue), and multiple interband transitions (orange). The data at ambient pressure were taken from a previous study\cite{Uykur2021b}. \textbf{e} and \textbf{f} Pressure evolution of the Drude and localization peak, respectively. \textbf{g} Calculated spectral weight of the two intraband contributions (Drude and localization peak). \textbf{h} Pressure dependence of the experimental plasma frequency deduced from the total intraband spectral weight as explained in the text and plasma frequency calculated by DFT. \textbf{i} Schematic pressure phase diagram of \CVS\ as determined by several electrical transport, magnetic susceptibility, NMR, XRD, and $\mu$SR studies\cite{Chen2021, Wang2021, Yu2021, Zhao2021, Zhu2022, Li2022, Chen2021b, Guguchia2022, Wang2021b, Du2021, Zhang2021, Zheng2022, Gupta2022}. The gray shaded areas mark critical pressure regions corresponding to (i) the vanishing of CDW state at $\sim$~2~GPa (ii) the disappearance of the first superconducting dome at around 9~GPa, and (iii) the re-emergence of superconductivity at approximately 12~GPa.}
\label{Fig2}
\end{figure*}

\begin{figure*}
\centering 
\includegraphics{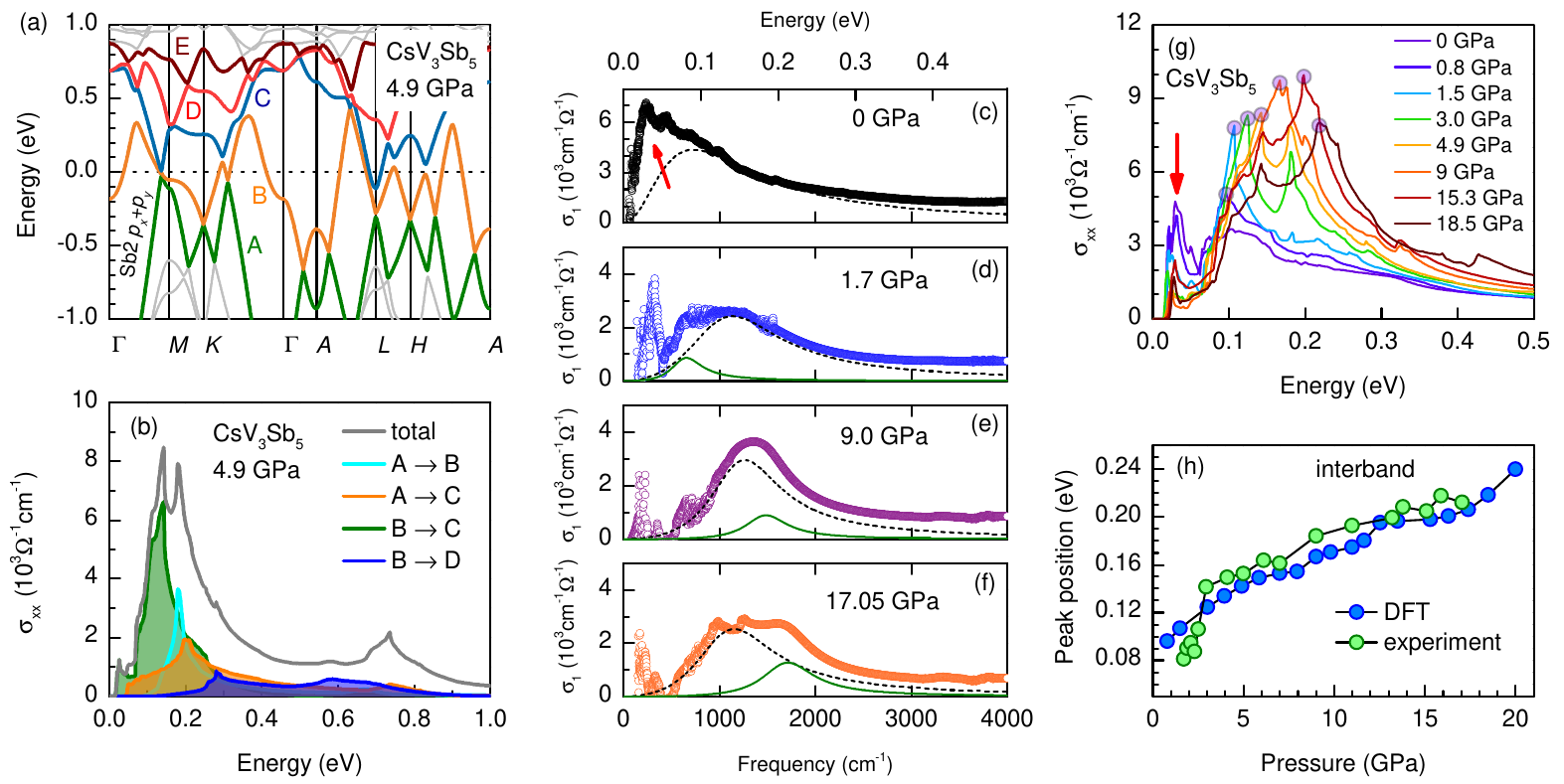}
\caption{\textbf{Pressure evolution of the band structure and interband transitions.} \textbf{a} Calculated band structure at 4.9~GPa. \textbf{b} Band-resolved contributions to the in-plane component of the calculated optical conductivity. \textbf{c-f} Experimental interband transitions at selected pressures. The red arrow marks the low-energy transitions ($\omega < 0.06$~eV) that become suppressed abruptly between 0 and 3~GPa. At low energies, the pressure data are described by two distinct absorption peaks (solid green line and black dotted line). With increasing pressure, a systematic blue shift of the green absorption is observed. \textbf{g} Pressure evolution of the calculated optical conductivity. The red arrow marks the
suppression of low-energy absorption and the circles highlight the shifting interband absorption peak. \textbf{h} Pressure evolution of the blue-shifting low-energy interband absorption peak from the experimental and calculated optical conductivities.}
\label{Fig3}
\end{figure*}

To further explore this interplay, we calculate the spectral weight (SW) by integrating the real part of the optical conductivity of the fitted Drude and localization peak according to\cite{Dressel2002}
\begin{equation}
\mathrm{SW} = \frac{1}{\pi^2\varepsilon_{\mathrm{0}} c}\int_{0}^{\omega_{\mathrm{c}}} \sigma_{\rm 1}(\omega) \rm d\omega,
\end{equation}
with $\varepsilon_{\mathrm{0}}$ being the permittivity in vacuum and $c$ the speed of light. The cut-off frequency is chosen as $\omega_{\mathrm{c}} = 50000$~\cm\ considering the high-energy tail of the localization peak. Fig.~\ref{Fig2}~(g) shows the pressure evolution of the Drude and localization peak spectral weight, as well as of the total spectral weight of the intraband processes (Drude + localization peak). While the total intraband SW is found to be almost unaffected by pressure, a redistribution of spectral weight between the Drude and the localization peak is observed.

The plasma frequencies are obtained via \wp~=~$\sqrt{\mathrm{SW_{Drude}} + \mathrm{SW_{loc}}}$ and normalized to the value at ambient pressure from a previous study\cite{Uykur2021b} (see Supplementary Note 3 for details). As displayed in Fig.~\ref{Fig2}~(h), a good match between the experimental and DFT plasma frequency is observed. Here, even small features in the experimental plasma frequency like dips around 2.5~GPa, 8~GPa, and 11~GPa are well reproduced by our calculations. This good agreement between experiment and \textit{ab initio} calculations over a wide pressure range suggests that electrons in \CVS\ are almost uncorrelated, consistent with the previous assessment of the correlation strength at ambient pressure\cite{Uykur2021b}. Hence, the main effect beyond the pure band picture is the damping of charge carriers manifested by the localization peak.

\textbf{Interband transitions.} Having discussed the intraband contributions in great detail, we now turn to analyzing the low-energy interband transitions. It has been shown that, compared to its sibling compounds, \CVS\ possesses inverted band saddle points, leading to very different low-energy interband transitions\cite{Wenzel2022, Uykur2021b}. At ambient pressure, optical spectroscopy has revealed the existence of characteristic interband absorptions below 0.06~eV (marked by the red arrow in Fig.~\ref{Fig3}~(c)), which could be well reproduced by DFT\cite{Uykur2021b}. 

\begin{figure*}
\centering 
\includegraphics{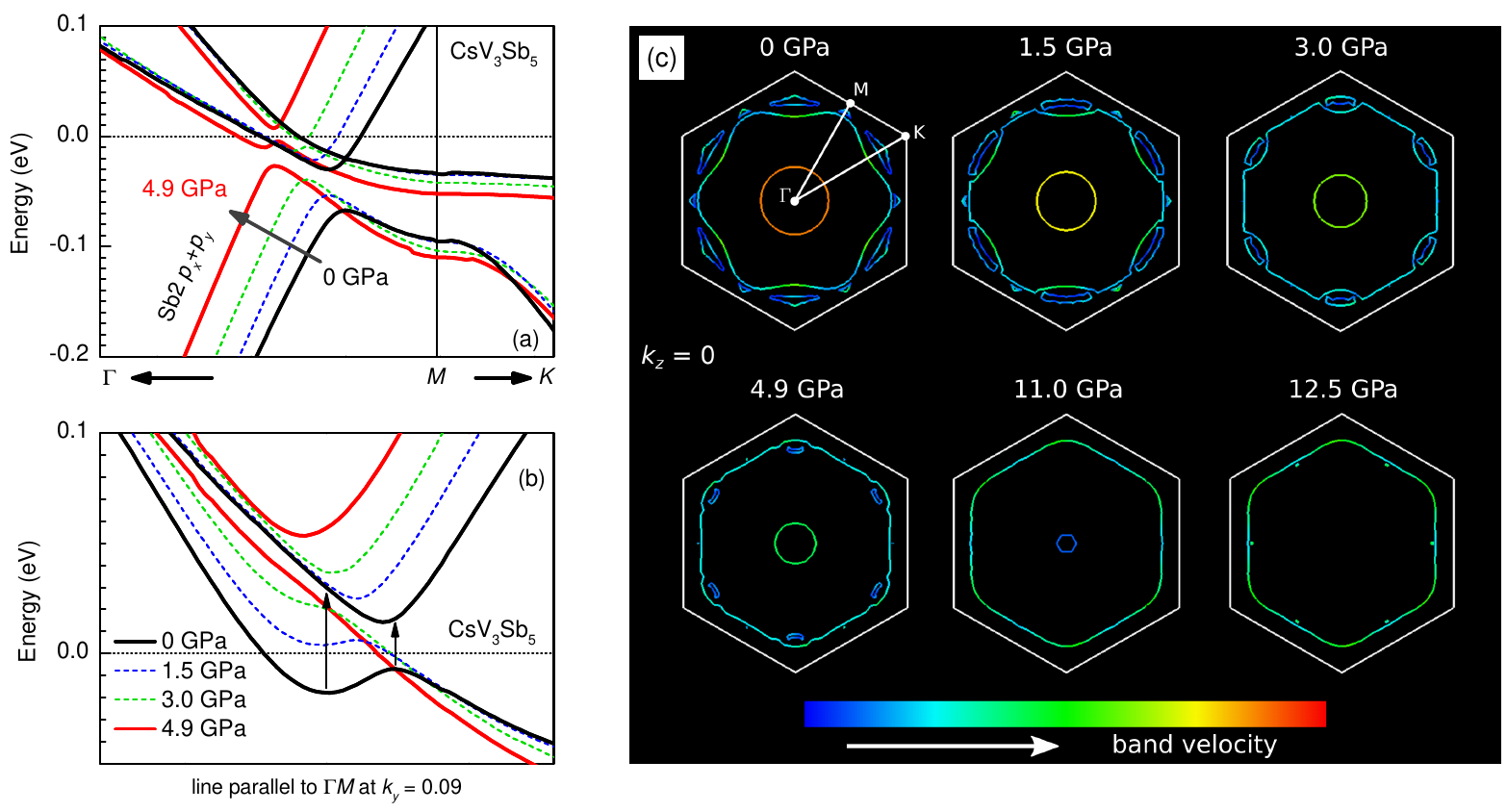}
\caption{\textbf{Low-pressure band structures and pressure evolution of the Fermi surface.} \textbf{a} Band structure along $\Gamma$~-~$M$ at low pressures. \textbf{b} Band structure away from the $\Gamma$~-~$M$ line shows the suppression of the low-energy interband transitions highlighted by black arrows. \textbf{c} Pressure evolution of the Fermi surface at $k_z = 0$, illustrated using \texttt{FermiSurfer}\cite{Kawamura2019} with the band velocity as color scale.}
\label{Fig4}
\end{figure*}

Upon applying pressure, the experimental data show a sudden reduction of the low-energy transitions at around 0.03~eV, already at the lowest measured pressure of 1.7~GPa as depicted in Fig.~\ref{Fig3}~(d). This trend is corroborated by the calculated optical conductivity plotted in Fig.~\ref{Fig3}~(g). Note that the difference in intensity between theoretical and experimental interband absorptions is due to the absence of electron scattering in the zero-temperature calculations. Band-resolved optical conductivity calculations displayed in Fig.~\ref{Fig3}~(b) for 4.9~GPa, where the spectral weight below 0.06~eV is almost suppressed, reveal that these low-energy absorption arises from transitions between bands B and C according to the labeling in Fig.~\ref{Fig3}~(a). A closer look at the knot of bands that cross the Fermi level along $\Gamma-M$ shown in Fig.~\ref{Fig4}~(a) suggests that this part of the band structure changes significantly already at low pressures. With the Sb2 $p_x + p_y$ band becoming steeper and pushing vanadium bands above the Fermi level, several bands crossing the Fermi level disappear, and parts of the Fermi surface shrink around 3~GPa, illustrated in Fig.~\ref{Fig4} (c). This effect is even better visible away from the $\Gamma$~-~$M$ line (Fig.~\ref{Fig4}~(b)), where interband transitions at 0.03~-~0.06~eV are clearly suppressed by pressure, and only interband transitions at $\omega > 0.1$~eV remain.

With this suppression of absorption below 0.1~eV, the experimental interband optical conductivity at low energies can be described by two absorption peaks as plotted in Figs.~\ref{Fig3}~(d-f). One of these peaks only slightly changes with pressure (black dotted line), while the other reveals a shift to higher energies (solid green line). This behavior is in line with our calculations showing a systematic shift of band C around the $M$-point away from the Fermi level, leading to a blue shift of the absorption peak related to the B-C transitions. The high consistency of the energy shift between calculations and experiment, as depicted in Fig.~\ref{Fig3}~(h), further supports the use of the uncorrelated band picture for the description of \CVS\ over a broad pressure range.  

Moreover, due to the increasing slope of the Sb2 $p_x + p_y$ band under pressure, band A comes closer to the Fermi level and even crosses it at higher pressures (see Supplementary Note 4). Consequently, the contribution of the A-B transitions increases upon applying pressure, compared to the B-C transitions that were dominant at ambient pressure, leading to an increase of spectral weight at around 0.2~eV (see Fig.~\ref{Fig3} (g)).

\section*{Discussion}
Previously, several optical studies demonstrated a close link between the low-energy interband absorptions and fine details in the electronic band structure of the \AVS\ series\cite{Uykur2021b, Uykur2022, Wenzel2022}. On the other hand, the localization peak, a common feature among several kagome metal compounds\cite{Uykur2021b, Uykur2022, Wenzel2022, Wenzel2022b, Biswas2020}, can not be explained within the simple band picture. Free charge carriers interacting with low-energy degrees of freedom, such as phonons, and electric or magnetic fluctuations, can lead to a backscattering of the electrons, causing localization effects manifested in a displaced Drude peak\cite{Fratini2014, Fratini2021}. In the absence of magnetism and electronic correlations, interactions between electrons and phonons become the most plausible reason for the appearance of the localization peak. This interpretation is further supported by the gradual shift of the localization peak to low energies on cooling, as phonons are suppressed\cite{Uykur2021b}. Moreover, optical studies of \KVS\ and \RVS\ find strong phonon anomalies, which are associated with the phonon modes coupling to the electronic background\cite{Wenzel2022, Uykur2022}. In the magnetic rare-earth kagome metal series $R$Mn$_6$Sn$_6$, a close link between the behavior of phonon modes and the localization peak was observed\cite{Wenzel2022b}.

It is then natural to use the spectral weight of the localization peak as a gauge of the electron-phonon coupling. Our data suggest that this coupling should be strongly suppressed around 2~GPa, where the CDW disappears. Concurrently, the increase in the spectral weight of the localization peak above 10~GPa indicates that the e-ph coupling becomes more prominent as superconductivity re-appears. On the microscopic level, first-principle calculations confirm this picture\cite{Si2022} and reveal a strong coupling of the V-V bond-stretching and V-Sb bond-bending phonon modes to the V 3$d_{xy,x^2-y^2,z^2}$, V 3$d_{xz,yz}$, and Sb1 5$p_z$ bands\cite{Wang2023}. Above 2~GPa, a drastic reduction of the coupling strength $\lambda$ is proposed, corresponding to the suppression of superconductivity. Consequently, above 12~GPa, the increase of the localization peak most probably signals an increase of e-ph coupling due to the new Fermi surface around $A$, as displayed in Fig.~\ref{Fig1}~(b)~and~(c). Here, Sb2 $p_z$ electrons appear at the Fermi level for the first time (see Supplementary Fig.~4 for the band structures at higher pressures), and probably cause the re-entrant superconductivity as illustrated in Fig.~\ref{Fig2}~(i). Given the unusual non-monotonic behavior of $T_{\mathrm{c}}$ under pressure, several other scenarios influencing superconductivity in \CVS\ can be excluded. Due to only weak changes in the plasma frequency (Fig.~\ref{Fig2}(h)), as well as a very different pressure evolution of the electronic density of states at $E_{\mathrm{F}}$ (see Supplementary Fig.~3~(d)), the behavior of $T_{\mathrm{c}}$ cannot be explained by modifications in the electronic structure. Moreover, the continuous evolution of the crystal structure\cite{Tsirlin2022, Tsirlin2022b, Yu2022b}, as well as of phonons according to previous DFT calculations\cite{Zhang2021b, Si2022}, eliminates the Debye temperature as a dominant factor, leaving changes in the e-ph coupling as the most likely reason for the observed pressure evolution of $T_{\mathrm{c}}$.

Apart from high-pressure investigations, \CVS\ was subject to various doping studies\cite{Xiao2023, Oey2022, Li2022b, Liu2022, Song2021, Zhou2023, Kato2022}. We would like to point out that while in some cases, for instance, through partial substitution of V-atoms with Nb-atoms, the effects on the electronic properties are similar to applying pressure, i.e., the suppression of the CDW order, along with an enhancement of \Tc, doping has a somewhat different effect on the band structure. As revealed by our study, the energies of the band saddle points are only weakly affected by pressure, while the Sb2 $p_x + p_y$ bands show a significant upward shift already at moderate pressures. On the other hand, Nb-doping leads to modifications near the $\Gamma$- and $M$-points, highly affecting the band saddle points \cite{Li2022b, Kato2022}. These differences in the band structure evolution entail a distinct behavior of the low-energy interband optical transitions. The application of external pressure leads to a systematic shift of band B away from the Fermi energy around the $M$-point, resulting in a blue-shifting interband absorption peak. Conversely, this low-energy absorption peak at around 400~\cm\ shifts to lower energies upon Nb-substitution\cite{Zhou2023}.

Our study revealed that the reduction in the e-ph coupling around 2~GPa is likely the main reason for the CDW suppression in \CVS. The simultaneous change in the interband absorption gives us a strong hint that this behavior is caused by the reduction of the FS along $\Gamma$~-~$M$. The Sb2~$p_x + p_y$ band, which was so far almost disregarded in the context of the \CVS\ physics, shows an upward shift and pushes some of the V~3$d$ bands above the Fermi level. This effect is qualitatively different from the scenario of band saddle points at $M$ moving away from the Fermi level and pulling V~3$d$ bands down in energy, consequently increasing the FS. We thus find a delicate interplay between electronic and lattice degrees of freedom in \CVS\ and identify a large tunability of the kagome bands via changes in the Sb sublattice.

\section*{Methods}
\textbf{Optical Measurements.} High-quality single crystals were grown and prepared according to ref.~\cite{Ortiz2019}. For the optical measurements, a freshly cleaved sample with a surface area of 150~$\mu$m~$\times$~150~$\mu$m and a thickness of $\sim$~60~$\mu$m was used.

High-pressure reflectivity measurements were performed at room temperature at the \textit{SMIS} beamline of the \textit{SOLEIL} synchrotron, France, on a homemade horizontal microscope with custom Schwarzschild objectives (NA~=~0.5). A diamond anvil cell (DAC) with type-IIa diamond anvils and a culet of 400 $\mu$m diameter was utilized. Finely ground CsI powder served as the pressure transmitting medium, making it possible to reach pressures up to 17.05~GPa. The sample and Ruby spheres used as pressure gauges were placed inside a stainless steel gasket with a 200~$\mu$m diameter hole. The pressure was determined by monitoring the calibrated shift of the ruby R1 fluorescence line as described in ref.~\cite{Mao1986}.

The reflectivity spectra at the sample-diamond interface were recorded in a broad spectral range of 150~-~10000~\cm\ by a Thermo-Fisher iS50 interferometer with KBr and solid substrate beamsplitters, using a MCT detector and a liquid helium-cooled bolometer. The reflectivity of a gold foil loaded into the DAC at ambient pressure served as a reference. Other optical quantities like the complex optical conductivity $\tilde{\sigma}(\omega)$, or the dielectric permittivity $\tilde{\varepsilon}(\omega)$, were obtained using standard Kramers-Kronig (KK) analysis considering the sample-diamond interface as explained in Supplementary Note 1.

\textbf{Computational Details.} Density-functional (DFT) band-structure calculations were performed with the \texttt{Wien2K} code\cite{Blaha2019, Blaha2020} and cross-checks have been conducted with the \texttt{FPLO} code\cite{Koepernik1999}. In all cases, the Perdew–Burke–Ernzerhof flavor of the exchange-correlation potential\cite{PBE1996} was used, and self-consistent calculations were converged on the $k$-mesh with 24~$\times$~24~$\times$~12 points. Experimental crystal structural parameters from refs.~\cite{Ortiz2019, Tsirlin2022} were used. Considering that the monoclinic distortion does not have a fundamental impact on the band structure\cite{Tsirlin2022b}, we used hexagonal structure throughout the pressure range for a simpler comparison.  Optical conductivity was calculated using the \texttt{optic} module\cite{Draxl2006} on the dense $k$-mesh with up to 100~$\times$~100~$\times$~50 points. Spin-orbit coupling was included in all the calculations. 

\textbf{Data availability.} The data that support the findings of this study are available from the corresponding authors upon request.

\section*{Acknowledgments}
We are grateful to Gabriele Untereiner for preparing the single crystals for the optical measurements. We thank SOLEIL synchrotron, France, for providing the beamtime (proposal~No.~20210399). M.W. is supported by IQST Stuttgart/Ulm via a project funded by Carl Zeiss foundation. S.D.W. and B.R.O. gratefully acknowledge support via the UC Santa Barbara NSF Quantum Foundry funded via the Q-AMASE-i program under award DMR-1906325. B.R.O. also acknowledges support from the California NanoSystems Institute through the Elings fellowship program. The work has been supported by the Deutsche Forschungsgemeinschaft (DFG) via DR228/51-3 and UY63/2-1. E.U. acknowledges the European Social Fund and the Baden-Württemberg Stiftung for the financial support of this research project by the Eliteprogramme. Computations for this work were done (in part) using resources of the Leipzig University Computing Center.

\section*{Additional information}

\textbf{Competing interests} The authors declare no competing financial or non-financial interests.\\\\
\textbf{Author contributions} M.W., E.U. and Y.T.C performed the experiments with technical support from F.C.. A.A.T. performed the DFT calculations. The samples were grown by B.R.O. and S.D.W. The manuscript was written by M.W., E.U., A.A.T, and M.D. with suggestions from all authors.\\\\
\textbf{Supplementary Information} accompanies this paper at\\\\
\textbf{Correspondence} Correspondence and requests for materials should be addressed to Ece Uykur (email: e.uykur@hzdr.de), Maxim Wenzel (email: maxim.wenzel@pi1.physik.uni-stuttgart.de), and Alexander A. Tsirlin (email: altsirlin@gmail.com).

\end{document}


\begin{center}
\textbf{Supplementary Information for\\''Pressure evolution of electron dynamics in the superconducting kagome metal CsV$_3$Sb$_5$''}

Maxim Wenzel, Alexander A. Tsirlin, Francesco Capitani, Yuk T. Chan, Brenden R. Ortiz, Stephen D. Wilson, Martin~Dressel, Ece Uykur

\end{center}

\textbf{This PDF file includes:}\\
Supplementary Note 1. Calculation of optical functions\\
Supplementary Note 2. Decomposition of optical spectra\\
Supplementary Note 3. Plasma frequency and electronic correlations\\
Supplementary Note 4. Additional computational results\\
Supplementary Figures 1-4\\
Supplementary References

\section*{Supplementary Note 1. Calculation of optical functions}
\label{KK}
In our study, the reflectivity at the sample-diamond interface $R_{\rm sd}(\omega)$ is measured at room temperature in a broad energy range (150-10000~\cm, equivalent to 18.6~meV-1.24~eV). Using a gold foil loaded into the diamond anvil cell as reference, the absolute value of the reflectivity is obtained as
\begin{equation}
R_{\rm sd}(\omega) = R_{\rm gold-d}(\omega) \frac{I_{\rm sd}(\omega)}{I_{\rm gold-d}(\omega)},
\end{equation}
with $I_{\rm sd}(\omega)$ and $I_{\rm gold-d}(\omega)$ denoting the intensity of the light reflected at the sample-diamond interface and the gold-diamond interface, respectively, and $R_{\rm gold-d}(\omega)$ being the reflectivity of the gold-diamond interface. The measured reflectivity at several pressures between 1.7 and 17.05~GPa is plotted in Supplementary Fig.~\ref{FigS1}~(a). For the extrapolations at low energies (green shaded area in the inset of Supplementary Fig.~\ref{FigS1}~(a)), required later for the Kramers-Kronig (KK) analysis, and in the range of strong diamond absorption (blue shaded area in the inset of Supplementary Fig.~\ref{FigS1}~(a)), the reflectivity is modeled by the Drude-Lorentz approach, with the complex dielectric function given by
\begin{equation}
\label{Eps}
\tilde{\varepsilon}(\omega)= \varepsilon_\infty - \frac{\omega^2_{\rm p,{\rm Drude}}}{\omega^2 + \mathrm{i}\omega/\tau_{\rm\, Drude}} + \sum\limits_j\frac{\Omega_{j}^2}{\omega_{\mathrm{0},j}^2 - \omega^2-\mathrm{i}\omega\gamma_j}.
\end{equation}
Here, $\varepsilon_{\infty}$ accounts for the high-energy contributions to the real part of the dielectric permittivity. $\omega_{\mathrm{p},{\rm Drude}}$ and $1/\tau_{\rm\,Drude}$ are the plasma frequency and the scattering rate of the itinerant carriers, respectively. The parameters $\omega_{\mathrm{0},j}$, $\Omega_j$, and $\gamma_j$ describe the resonance frequency, the width, and the strength of the $j^{th}$ excitation, respectively.  

The optical conductivity and dielectric function are calculated using Kramers-Kronig (KK) analysis. Here, the sample-diamond interface leads to an additional phase shift due to the refractive index of diamond being 2.38, which needs to be accounted for via the additional term in the brackets of Eq.~\ref{KKeq}\cite{Pashkin2006}
\begin{equation}
\label{KKeq}
\theta(\omega') = -\frac{\omega'}{\pi}P\int_0^{\infty}\frac{\mathrm{ln}R_{\rm sd}(\omega)}{\omega^2 - \omega'^2} \mathrm{d}\omega + \left[\pi - 2\mathrm{arctan}\frac{\omega_{\beta}}{\omega'}\right].
\end{equation} 
The value of $\omega_{\beta}$ is estimated by comparing the optical conductivity of the lowest measured pressure (1.7~GPa) calculated by KK analysis with the optical conductivity at ambient pressure. The best match at high frequencies is found for $\omega_{\beta} = 20000$~\cm. The calculated real part of the optical conductivity and the dielectric function are presented in Supplementary Fig.~\ref{FigS3} (b) and (c), respectively.

\begin{figure*}
\centering 
\includegraphics[width=\textwidth]{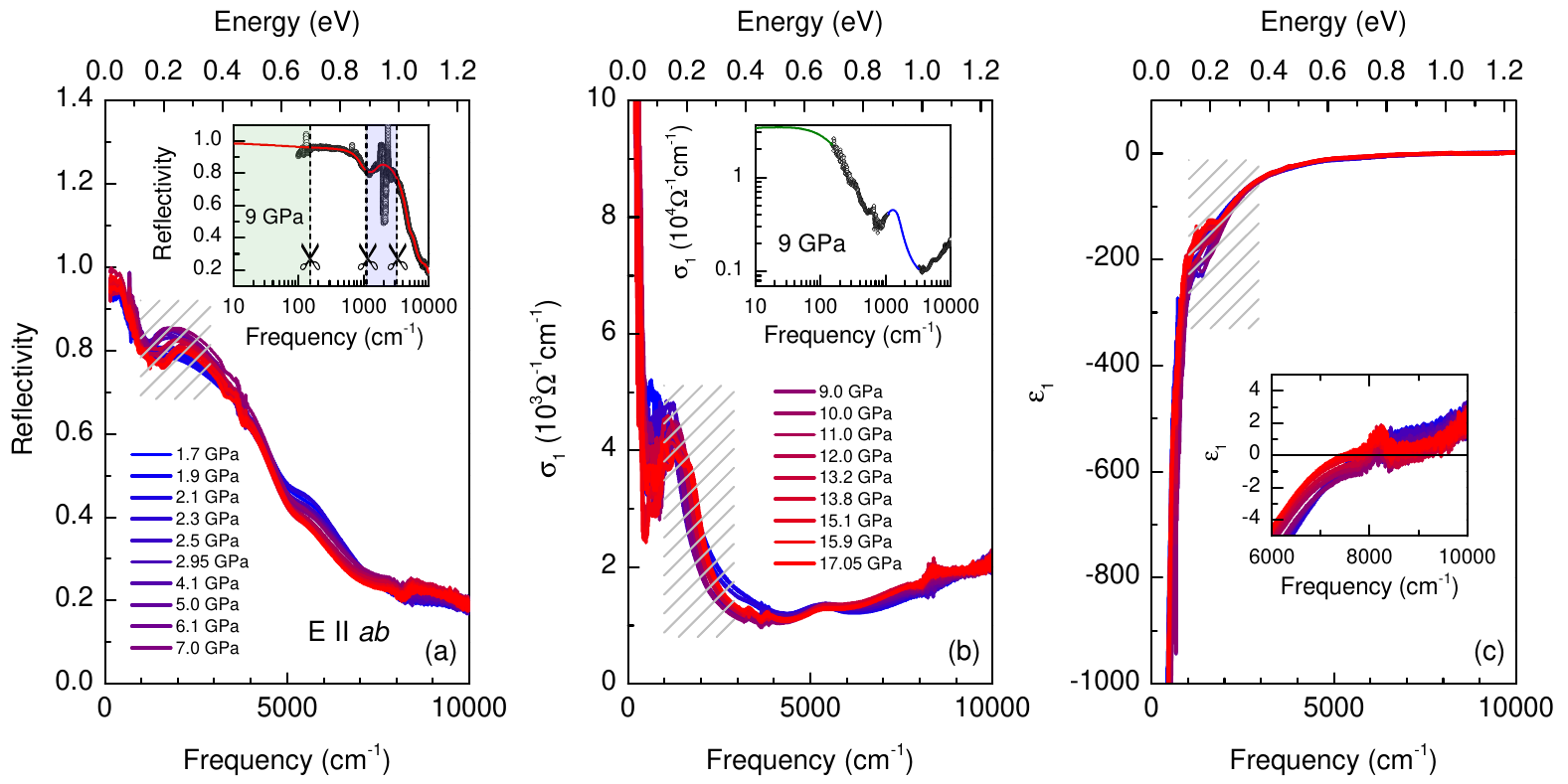}
\caption{\textbf{Reflectivity, optical conductivity, and dielectric permittivity.} \textbf{a} Absolute reflectivity at the sample-diamond interface, obtained as described in the text. The grey dashed lines mark the area of diamond phonon absorptions in the spectra. The inset shows the fit of the measured reflectivity exemplary at 9~GPa used to extrapolate the data to $\omega = 0$~eV (green shaded area) and to replace the data in the range of the strong phonon absorptions (blue shaded area). \textbf{b} Calculated real part of the optical conductivity. \textbf{c} Calculated dielectric permittivity. The inset displays the zero-crossing corresponding to the screened plasma frequency of the system with $\omega_{\mathrm{p}}^{\rm screened} = \omega_{\mathrm{p}} / \sqrt{\varepsilon_{\infty}}$~\cite{Dressel2002}.}
\label{FigS1}
\end{figure*}

\section*{Supplementary Note 2. Decomposition of optical spectra}
Several contributions to the optical conductivity are modeled using the Drude-Lorentz approach, as presented in Supplementary Note 1. The reflectivity and the complex optical conductivity are fitted simultaneously to ensure the reliability of the fit parameters. Interband absorptions are modeled employing a total of six Lorentzians. One single Drude component is not sufficient to describe the low-energy dynamics, but an additional broad peak at low energies associated with localized charge carriers is required for a satisfactory description. While, in principle, this peak can be fitted with a simple Lorentzian, we follow the approach of previous optical studies of kagome metals using a displaced Drude model proposed by Fratini \textit{et al.}\cite{Fratini2014}. Assuming interactions of charge carriers with low-energy excitations, such as phonons, electric or magnetic fluctuations leading to a backscattering of the electrons, this model gives physical reasoning for the observed localization. With $C$ being a constant, $\hbar$ the reduced Planck constant, $k_{\mathrm{B}}$ the Boltzmann constant, $\tau_{\mathrm{b}}$ the backscattering time, and $\tau$ the elastic scattering time of the standard Drude model, the real part of the optical conductivity reads\cite{Fratini2014}
\begin{equation}
\tilde{\sigma}_{\rm localization}(\omega)=\frac{C}{\tau_{\mathrm{b}}-\tau}\frac{\tanh\{\frac{\hbar\omega}{2k_{\mathrm{B}}T}\}}{\hbar\omega}\,\cdot\,\mathrm{Re}\left\{\frac{1}{1-\mathrm{i}\omega\tau}-\frac{1}{1-\mathrm{i}\omega\tau_{\mathrm{b}}}\right\}.
\end{equation}
The total dielectric permittivity takes the form
\begin{equation}
\tilde{\varepsilon}(\omega) = \tilde{\varepsilon}_{\rm Drude}(\omega) + \tilde{\varepsilon}_{\rm Lorentz}(\omega)+\tilde{\varepsilon}_{\rm localization}(\omega).
\end{equation}
The complex optical conductivity [$\tilde{\sigma}=\sigma_{\rm 1} + \mathrm{i}\sigma_{\rm 2}$] is then calculated as
\begin{equation}
\tilde{\sigma}(\omega)= -\mathrm{i}\omega\varepsilon_{\mathrm{0}}[\tilde{\varepsilon} (\omega) -\varepsilon_\infty].
\label{Cond}
\end{equation}
The fit parameters of the Drude and localization peak are plotted in Supplementary Fig.~\ref{FigS2} as a function of pressure. The behavior of the dc conductivity $\sigma_{\rm 0}$ is consistent with previous transport measurements\cite{Zhang2021}. Panels (g) and (h) illustrate the fits of the localization peak at various pressures. 
\begin{figure*}
\centering 
\includegraphics[width=\textwidth]{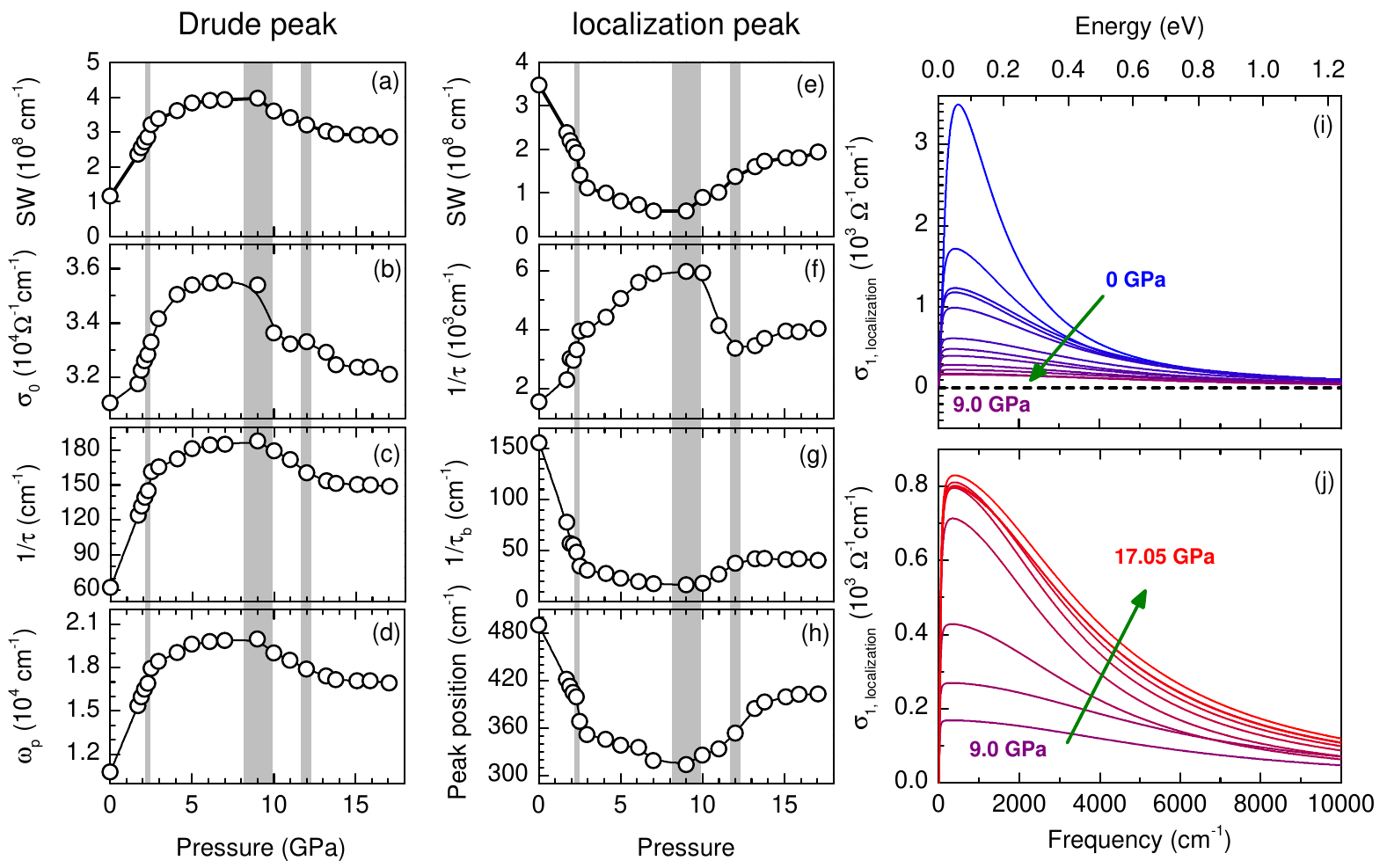}
\caption{\textbf{Fit parameters of the Drude and localization peaks.} \textbf{a-d} Spectral weight, dc conductivity, scattering rate, and calculated plasma frequency, respectively, of the Drude component. \textbf{e-h} Spectral weight, scattering rate, backscattering rate, and peak position, respectively, from fits with the Fratini model. The gray shaded areas mark critical pressure regions corresponding to (i) the vanishing of CDW state at $\sim$~2~GPa (ii) the disappearance of the first superconducting dome at around 9~GPa, and (iii) the re-emergence of superconductivity at approximately 12~GPa. \textbf{i} and \textbf{j}~Pressure evolution of the localization peak.}
\label{FigS2}
\end{figure*}

\section*{Supplementary Note 3. Plasma frequency and electronic correlations}
The plasma frequency is calculated via the spectral weight analysis of the Drude and localization peak presented in the main text. However, evaluating the exact Drude contribution is difficult, as only the tail is visible ($\omega > 150$~\cm). Moreover, the low-energy interband absorptions may affect the weight of the localization peak. Alternatively, the experimental plasma frequency can be calculated from the zero-crossing of the dielectric permittivity corresponding to the screened plasma frequency $\omega_{\rm p}^{\rm screened} = \omega_{\rm p} / \sqrt{\varepsilon_{\infty}}$~\cite{Dressel2002}. Using this method, $\omega_{\rm p} = 3.13$~eV was determined at ambient pressure\cite{Uykur2021b}. Due to the limited energy range, it is not possible to pinpoint the value of $\varepsilon_{\infty}$ in our high-pressure study. Nevertheless, as displayed in Supplementary Fig.~\ref{FigS3}~(b), the zero-crossing of $\varepsilon_{\rm 1}$ is almost unaffected by pressure in line with the plasma frequency estimated from the spectral weight analysis. Hence, for consistency, we normalize the plasma frequency to $\omega_{\rm p} = 3.13$~eV at 0~GPa as shown in Supplementary Fig.~\ref{FigS3}~(a). The strength of electronic correlations can be assessed by the ratio $\omega_{\mathrm{p}, \mathrm{experiment}}^2 / \omega_{\mathrm{p}, \mathrm{DFT}}^2$\cite{Shao2020} signaling very weak and pressure-independent electronic correlations as depicted in Supplementary Fig.~\ref{FigS3}~(c).
\begin{figure*}
\centering 
\includegraphics[width=\textwidth]{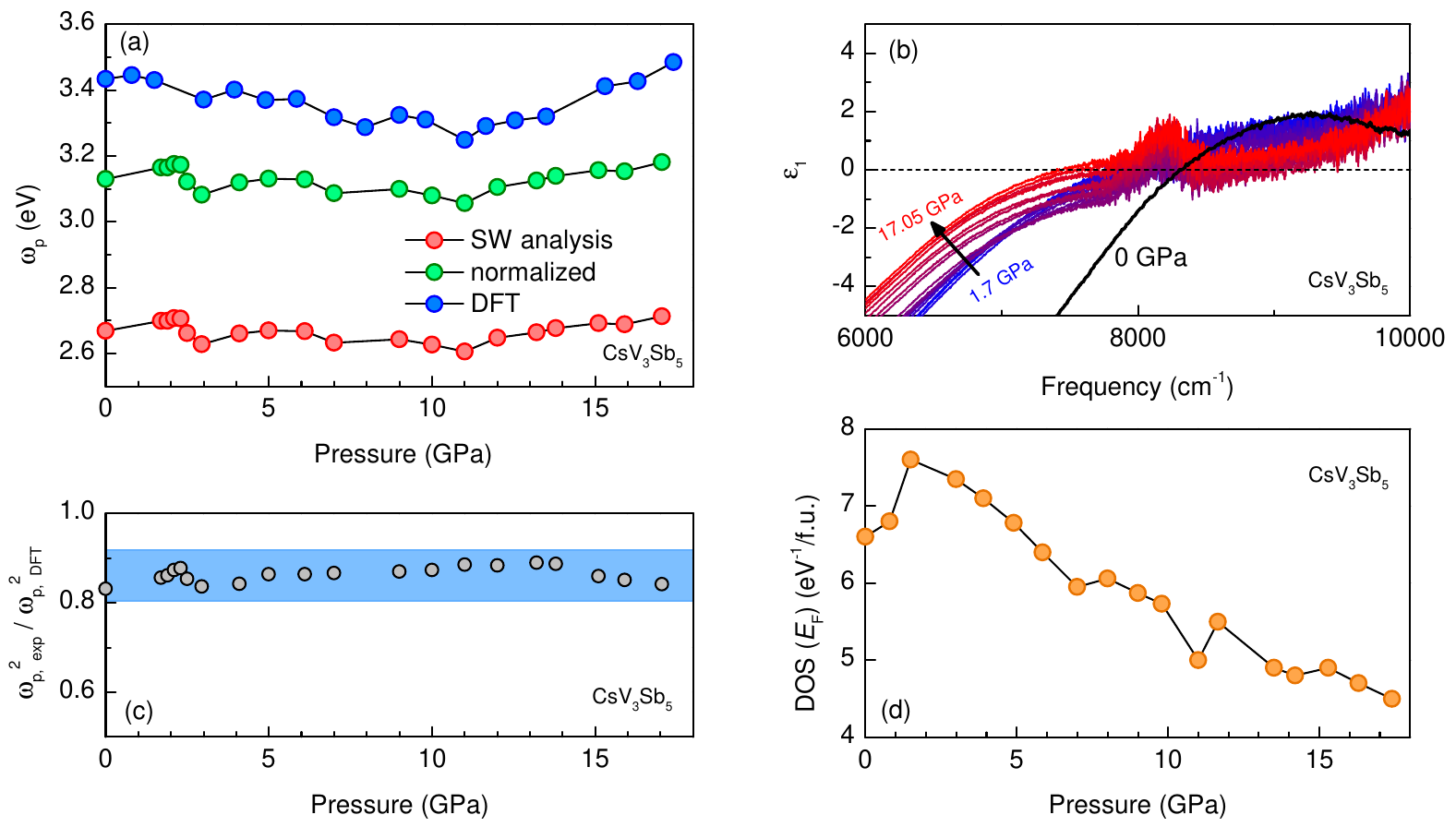}
\caption{\textbf{Plasma frequency, electronic correlations, and electronic density of states.} \textbf{a} Pressure evolution of the experimental plasma frequency determined from the spectral weight analysis (red) and DFT plasma frequency (blue). For consistency, we normalize the experimental plasma frequencies to the value at ambient pressure estimated from the zero-crossing of $\varepsilon_{\rm 1}$ as explained in the text (green). \textbf{b} Zero-crossing of the dielectric permittivity. \textbf{c}~Strength of electronic correlations as a function of pressure estimated from the ratio of the squared experimental and DFT plasma frequencies. This ratio stays close to 0.85 over the whole measured pressure range, as illustrated by the blue-shaded area, suggesting weak and pressure-independent electronic correlations. \textbf{d} Electronic density of states at the Fermi energy as a function of pressure.}
\label{FigS3}
\end{figure*}

\section*{Supplementary Note 4. Additional computational results}
As discussed in the main text, several scenarios leading to the unusual non-monotonic behavior of $T_{\mathrm{c}}$ can be envisaged. In Supplementary Fig.~\ref{FigS3}~(d), we demonstrate that changes in the electronic density of states cannot be responsible for the peculiar double-dome pressure evolution of superconductivity.

\begin{figure*}[h]
\centering 
\includegraphics[width=\textwidth]{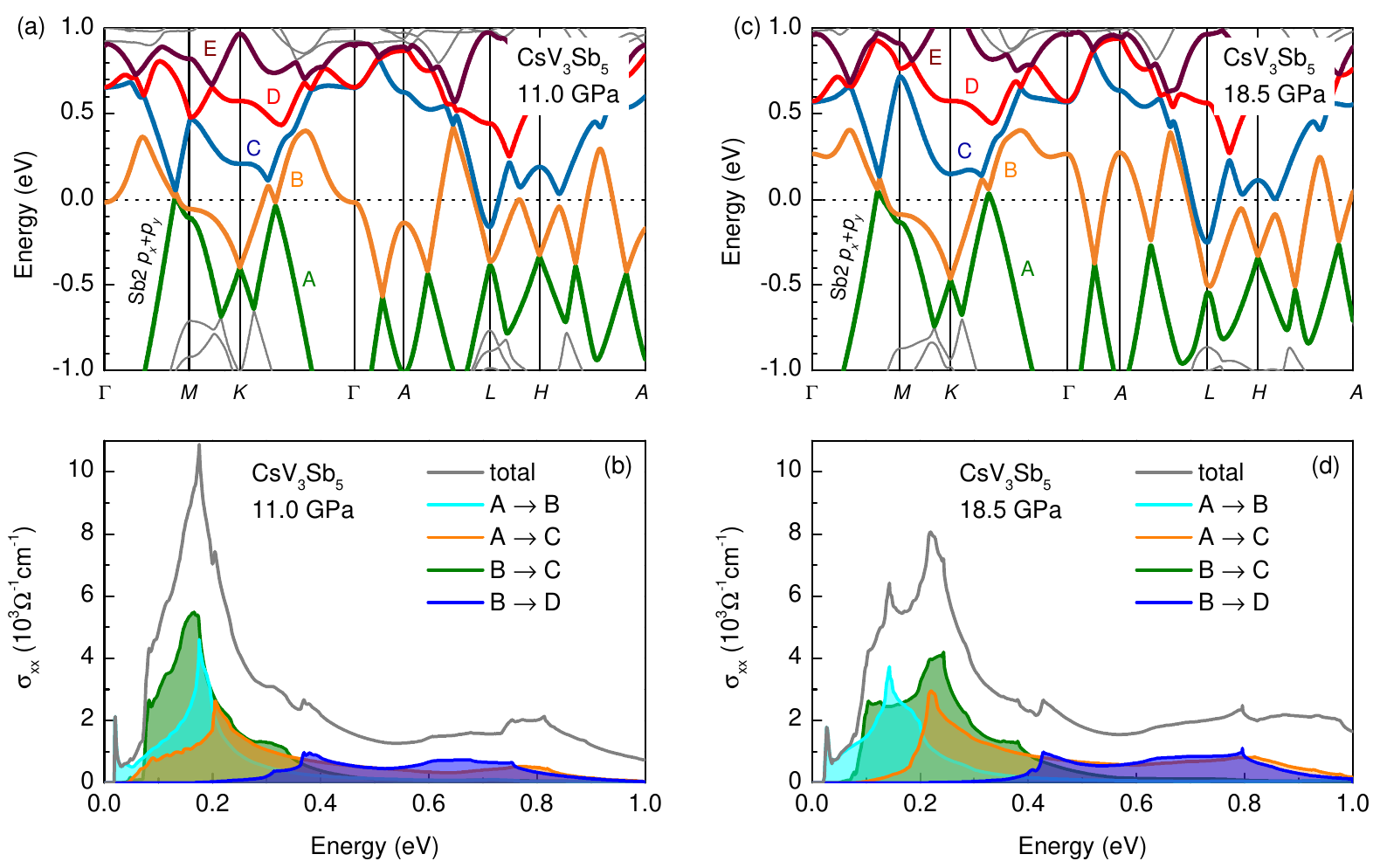}
\caption{\textbf{Band structure and band-resolved optical conductivity.} \textbf{a} and \textbf{c} Calculated band structures at 11 and 18.5~GPa, respectively. \textbf{b} and \textbf{d} Band-resolved in-plane optical conductivity at 11 and 18.5~GPa, respectively.}
\label{FigS4}
\end{figure*}

Supplementary Figs.~\ref{FigS4} (a) and (c) show the calculated band structures at 11 and 18.5~GPa, respectively. The close comparison reveals two main pressure-induced modifications. (i) Around the $M$-point, band C systematically shifts to higher energies, and (ii) Band B comes closer to the Fermi level at the $A$-point and crosses $E_{\mathrm{F}}$ above 12~GPa. The appearance of these Sb2 $p_z$ electrons at the Fermi level correlates with the re-entrant superconductivity and the increase of the localization peak, signaling a gain in electron-phonon coupling. The calculated band-resolved optical conductivities are given in Supplementary  Figs.~\ref{FigS4} (b) and (d). While most of the main contributions are not crucially affected by pressure, transitions between bands B and C change drastically, leading to a blue shift of the absorption peak.

\section*{Supplementary References}